\documentclass[pre,aps,epsfig,twocolumn,amsmath,showpacs]{revtex4}
\usepackage{graphicx}
\usepackage{amssymb}
\usepackage{bm}

\begin{document}

\title{Conserved mass aggregation model with mass-dependent fragmentation}

\author{ Dong-Jin \surname{Lee}, Sungchul \surname{Kwon}, and Yup \surname{Kim}}
\email{ykim@khu.ac.kr  } \affiliation{Department of Physics and
Research Institute for Basic Sciences, Kyung Hee University, Seoul
130-701, Korea}

\date{\today}

\begin{abstract}
We study a conserved mass aggregation model with mass-dependent
fragmentation in one dimension. In the model, the whole mass $m$
of a site isotropically diffuse with unit rate. With rate
$\omega$, a mass $m^{\lambda}$ is fragmented from the site and
moves to a randomly selected nearest neighbor site. Since the
fragmented mass is smaller than the whole mass $m$ of a site for
$\lambda < 1$, the on-site attractive interaction exists for the
case. For $\lambda = 0$, the model is known to undergo the
condensation phase transitions from a fluid phase into a condensed
phase as the density of total masses ($\rho$) increases beyond a
critical density $\rho_c$. For $0< \lambda <1$, we numerically
confirm for several values of $\omega$ that $\rho_c$ diverges with
the system size $L$. Hence in thermodynamic limit, the condensed
phase disappears and no transitions take place in one dimension.
We also explain that there are no transitions in any dimensions.

\end{abstract}

\pacs{68.35.Ct,02.50.-r,05.40.a,64.60.Ht}

 \maketitle

Nonequilibrium phase transitions have been studied extensively
during last decades and observed in various systems
\cite{Liggett,Marro,hin,jkps}. As in equilibrium, nonequilibrium
transitions can be classified into universality classes according
to conservation laws and symmetries, which are characterized by
several critical exponents.

In this paper, we study condensation phase transitions.
Nonequilibrium condensation phase transitions from fluid phase
into condensed phase have been observed in a variety of phenomena
ranging from traffic flow to polymer gels
\cite{zrp,E2,LEC,M,noh,Z,W,S,MRRGI,F,MKB}. In the steady state, a
finite fraction of total particles condenses on a single site in
condensed phase when the total particle density $\rho$ is
increased beyond a certain critical value $\rho_c$. In fluid phase
below $\rho_c$, the particle number of each site fluctuates around
$\rho$ without the condensation \cite{zrp,MKB}.

The simplest model exhibiting the condensation is zero-range
process (ZRP) \cite{zrp}. In ZRP, many identical particles occupy
sites on a lattice. Each site may contain an integer number of
particles and one of these particles can hop to one of the nearest
neighboring sites with a rate that depends on the number of
particles on the site of departure. The chipping (single-particle
dissociation) and aggregation processes of ZRP describe various
condensations such as traffic jam \cite{E2}, bunching of buses
\cite{LEC}, coalescence of shaken steel balls \cite{M} and the
condensation on complex networks \cite{noh}.

Another important class of condensation transitions emerges when
the diffusion of the whole particles of a site is allowed in
addition to the chipping and aggregation. These processes arise in
a variety of phenomena such as polymer gels \cite{Z}, the
formation of colloidal suspensions \cite{W}, river networks
\cite{S,MRRGI} and cloud formation \cite{F}. Recently studied
symmetric conserved-mass aggregation(SCA) model is the simplest
one incorporating diffusion, chipping and aggregation upon contact
\cite{MKB,exactsca}. In one dimensional SCA model, the mass $m_i$
of site $i$ moves either to site $i-1$ or to site $i+1$ with the
unit rate, and then $m_i \rightarrow 0$ and $m_{i\pm1} \rightarrow
m_{i\pm1}+m_i$. With the rate $\omega$, unit mass chips off from
site $i$ and moves to one of the nearest neighboring sites; $m_i
\rightarrow m_i -1$ and $m_{i\pm1}\rightarrow m_{i\pm1}+1$. As
total masses are conserved, the conserved density $\rho$ and
$\omega$ determine the phase of SCA model. The condensation
transition arises from the competition between diffusion and
chipping processes. The diffusion of masses tends to produce
massive aggregates and consequently creates more vacant sites. The
chipping of unit mass tends to prevent the formation of
aggregates, so that it leads to a replenishment of the lower end
of mass distribution.

The single site mass distribution $P(m)$, i.e., the probability
that a site has mass $m$ in the steady state, was shown to undergo
phase transitions on regular lattices \cite{MKB}. For a fixed
$\omega$, as $\rho$ is varied across the critical value $\rho_c
(\omega)$, the behavior of $P(m)$ for large $m$ was found to be
\cite{MKB}

\begin{equation}
 P(m)\sim
 \begin{cases}
          e^{-m/m^{*}}& \rho<\rho_c (\omega) \\
          m^{-\tau}& \rho=\rho_c (\omega) \\
          m^{-\tau}+ \text{\rm infinite\,\,\, aggregate}& \rho > \rho_c
          (\omega),
        \end{cases}
\end{equation}
where $\rho_c$ is given as $\rho_c (\omega) = \sqrt{\omega+1}-1$.
$\rho_c$ and $\tau$ are shown to be independent of the spatial
dimension $d$ \cite{exactsca}. The tail of the mass distribution
changes from exponential to an algebraic decay as $\rho$
approaches $\rho_c$ from below. As one further increases $\rho$
beyond $\rho_c$, this asymptotic algebraic part of the critical
distribution remains unchanged, but in addition an infinite
aggregate forms. This means that all the additional mass
$(\rho-\rho_c)L^d $ condenses onto a single site and does not
disturb the background critical distribution. The $\omega =
\infty$ case corresponds to ZRP with a constant chipping rate, and
then there is no condensation transitions on regular lattices
\cite{zrp}. The critical exponent $\tau$ is same everywhere on the
critical line $\rho_c (\omega)$. Recent studies showed that the
existence of the condensation transitions depends on the spatial
disorder \cite{jain}, the symmetry of moving directions \cite{RK},
the constraint of diffusion rate \cite{ca-mass}, and the
underlying network structure \cite{cma-network}.

In more general situations, as pointed out in Ref. \cite{MKB}, the
diffusion rate of mass and the chipped mass will depend on the
mass of a departure site. When the diffusion rate depends on mass
as $D(m) \sim m^{-\alpha}$ with $\alpha >0$ \cite{ca-mass}, the
condensed phase disappeared in the thermodynamic limit. However
finite systems undergo the condensation transitions at a certain
critical density $\rho_c$ which algebraically diverges with the
system size $L$. On the other hand, the mass-dependent
fragmentation or chipping is also important in general physical
situation. For instance, in gelation phenomena, there is no reason
that only one monomer at the end of a polymer is chipped
regardless of the mass of polymers. It is very likely that more
monomers can be chipped for more massive polymers. Hence it is
natural to study the effect of mass-dependent fragmentation on the
condensation phase transitions. In this paper, we investigate the
condensation transitions of SCA models in which the fragmented
mass in a chipping process depends on the whole mass of the
departure site.

We consider SCA model with mass-dependent fragmentation in one
dimension. In the model, the fragmented mass in the chipping
process is given as $\delta m = A m^\lambda$. By chipping process
with the rate $\omega$, the mass $\delta m$ randomly moves to the
one of the nearest neighbors, $m_i \to m_i - \delta m_i$ and
$m_{i\pm1} \rightarrow m_{i\pm1}+\delta m_i$. With the unit rate,
mass $m_i$ randomly diffuses to the one of the nearest neighbors,
$m_i \rightarrow 0$ and $m_{i\pm1} \rightarrow m_{i\pm1}+m_i$.

Since the chipped mass is smaller than the whole mass of a site
for $\lambda <1$, the attractive interaction exists between masses
on the same site. The $\lambda = 0 $ case is just SCA model of
Ref. \cite{MKB}. The $\lambda = 1$ is a trivial point, but two
extreme situations emerge according to the values of $A$. For
$A=0$ or $1$, masses only diffuse without chipping so the complete
condensation takes place. On the other hand, for $0<A<1$, the
fragmented mass is proportional to the whole mass, which means
that any aggregates cannot exit. Hence, the system is always in
the fluid phase. The $\lambda=1$ and $0<A<1$ case corresponds to
the random fragmentation model in which $P(m)$ always
exponentially decays \cite{rf}. On the other hand, for $\lambda
<0$, the complete condensation should always occur because the
time needed to dissipate an aggregate, if exists, exponentially
increases with the mass of an aggregate. Therefore, we focus on
the range of $0<\lambda<1$.

We perform Monte Carlo simulations on a ring of size $L$. With
random initial distributions of total mass $\rho L$, the system is
allowed to reach the steady states. We run simulations typically
up to $t=10^9$ time steps for size $L$ up to $2048$. In the steady
states, we measure the single-site mass distribution $P(m)$ for
various values of $\lambda$ and $\omega$.
\begin{figure}
\includegraphics[width=8cm]{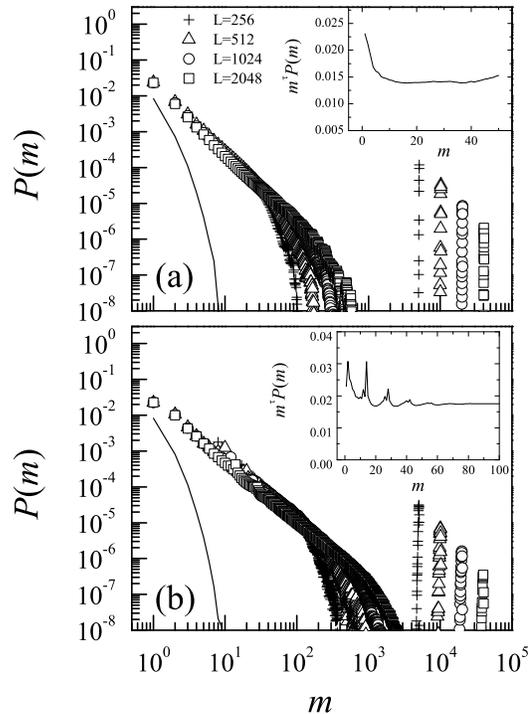}
\caption{ The double logarithmic plots of $P(m)$ for $\lambda = 0.1$
(a) and $0.25$ (b) with $\omega=0.1$. Insets shows the scaling plot
$m^{-\tau}P(m)$ with $\tau = 1.84$ (a) and $1.62$ (b) for the data
of $L=2048$. In each panel, the solid line corresponds to
$\rho=0.01$ with $L=2048$. Symbols are the data of $\rho = 20$ with
$L$ up to $2048$. } \label{fig.1}
\end{figure}
Fig.1 shows $P(m)$ of $\lambda=0.1$ and $0.25$ with $\rho=20$ and
$\omega = 0.1$ for $L$ up to $2048$. For both $\lambda$ values, an
aggregate with mass $(\rho - \rho_c)L$ forms and its probability
decreases as $1/L$. $\rho_c$ is a certain critical density above
which an aggregate forms. $P(m)$ shows that in finite systems, the
condensation phase transitions occurs at finite $\rho_c$ for
$\lambda >0$. For other values of $\omega$, the condensation
transitions also occur in finite systems. $\rho_c$ increases with
$\omega$ while $\tau$ is the same for all $\omega$.

If there exists the condensation transitions in the thermodynamic
limit, $\tau$ should satisfy $\tau >2$. We measure the exponent
$\tau$ from the scaling plot of $m^{\tau}P(m)$. For $L=2048$ with
$\omega=0.1$, we estimate $\tau$ as $1.84(10)$ for $\lambda=0.1$ and
$1.62(10)$ for $\lambda=0.25$. For other values of $\omega$, we
obtain the similar $\tau$ within errors. However, the finite size
effect appears at very small mass even for large $L$, for example,
at about $m=100$ for $L=2048$. Even though the precise measurement
of $\tau$ is difficult due to the strong finite size effect, our
estimates of $\tau$ are smaller than $2$. It means that the
condensation transitions take place only in the finite systems and
disappear in the thermodynamic limit. In what follows, we discuss
that no condensation transitions occurs in the thermodynamic limit.

\begin{figure}
\includegraphics[width=8cm]{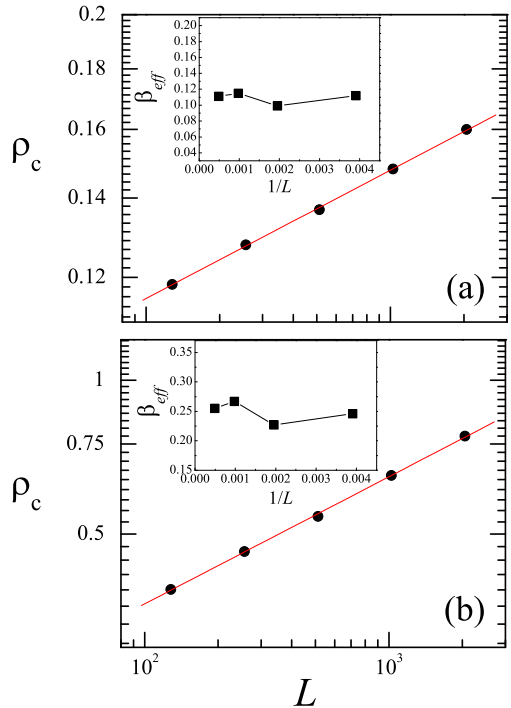}
\caption{ The double logarithmic plots of $\rho_c (L)$ for
$\lambda = 0.1$ (a) and $0.25$ (b) with $\omega=0.1$ and
$\rho=20$. The slope of line in each panel is $\beta = 0.1$ (a)
and $0.25$ (b). } \label{fig.1}
\end{figure}

In infinite systems with finite density ( or $L \to \infty$ ),
$\tau$ must be greater than $2$ since the first moment $<m>$ is
the density. Otherwise, the first moment diverges as $\ln L$ for
$\tau=2$ and $L^{\beta}$ for $\tau <2$ respectively. Hence $\tau <
2$ for $\lambda >0$ contradict with the conservation of total
density. Similar finite size effect with $\tau \leq 2$ has been
observed in asymmetric CA \cite{RK} and SCA with mass-dependent
diffusion rate \cite{ca-mass}. $\rho_c$ diverges with $L$ as
$\rho_c \sim \ln L$ for asymmetric CA and $L^\beta$ for the
mass-dependent diffusion rate respectively. In the thermodynamic
limit or $L \to \infty$, no transitions occurs due to $\rho_c =
\infty$ and a system is always in the fluid phase. As a result,
$\tau <2$ for $\lambda>0$ implies that $\rho_c = \infty$ and the
condensation transitions disappear in the thermodynamic limit.
Only finite systems undergo the condensation transitions at a
finite $\rho_c$.

To find the size dependence of $\rho_c$, we measure $\rho_c$ for
each $L$ as follows. In the steady states, $P(m)$ scales as
\cite{exactsca}

\begin{equation}
P(m)=m^{-\tau} f(m/L^{\phi})+\frac{1}{L}\delta (m-(\rho-\rho_c)L)
\;,
\end{equation}
where $\phi$ is the crossover exponent \cite{exactsca}. From the
conservation of total masses, $\rho_c$ is given as $\rho_c = \rho
- \rho_{\infty}$, where $\rho_\infty$ is the density of an
aggregate. From the fact that the background distribution does not
change for $\rho \geq \rho_c$, one can estimate $\rho_c$ via
$\int_{1}^{m_0} m P(m) dm$ in the condensed phase, where $m_0$ is
the cut-off mass at which the background distribution terminates.
Using this method, we confirmed that our estimate for $\rho_c$ of
the normal SCA model \cite{MKB} converges to the the exact
$\rho_c$ as $L$ increases (not shown). In Fig. 2, we plot $\rho_c$
as a function of $L$ for $\lambda=0.1$ and $0.25$ with
$\omega=0.1$, where we measure $\rho_c$ using $P(m)$ for $\rho =
20$. Insets show the local slope $\beta_{eff}$ defined as
$\beta_{eff}=\ln [\rho(L)/\rho(L/b)]/\ln b$ with $b=2$. As one can
see, $\rho_c$ diverges with $L$ as $L^{\beta}$ with
$\beta=0.11(3)$ for $\lambda=0.1$ and $\beta = 0.25(5)$ for
$\lambda=0.25$. Since the fragmented masses increases for $\lambda
\to 1$, $\rho_c$ should diverge faster for larger $\lambda$. Hence
we conclude that there are no condensation transitions in the
thermodynamic limit for $0<\lambda <1$.

The scaling relation between three exponents, $\tau$, $\phi$ and
$\beta$ is obtained by calculating $\rho_c = \int_{1}^{m_c} m P(m)
dm$ for $\rho \geq \rho_c$, where $m_c$ is the crossover mass over
which $P(m)$ decays exponentially due to the finite size effect.
The lower limit is the mass from which $P(m)$ begins to decay
algebraically to the upper limit $m_c$. We set the lower limit of
the integral to unit for convenience. Since $m_c$ scales with $L$
as $m_c \sim L^{\phi}$, $\rho_c$ scales as $L^{\phi(2-\tau)}$.
Hence $\beta$ is given as $\beta = \phi (2-\tau)$. From the
normalization condition of $P(m)$, one obtains another scaling
relation $\phi(\tau-1) = 1$ \cite{exactsca}. Hence we have the two
scaling relations,

\begin{equation}
\beta = \phi (2-\tau) \;\;, \;\; \phi(\tau-1) = 1 \;.
\end{equation}
From the two relations, $\tau$ is given as $\tau =
(\beta+2)/(\beta+1)$. With the estimates of $\beta$, we obtain
$\tau = 1.91$ for $\lambda=0.1$ and $1.8$ for $\lambda=0.25$
respectively. Since the direct measurement of $\tau$ crucially
depends on the choice of the scaling region, the estimate of
$\tau$ is very subjective. On the other hand, $\rho_c$ exhibits
the nice scaling with $L$ as shown in Fig. 2, the prediction of
$\tau$ by the scaling relation should be more precise. Hence we
predict $\tau = 1.91(3)$ for $\lambda=0.1$ and $1.80(3)$ for
$\lambda = 0.25$.

Recent studies showed that SCA models is well described by mean
field theory \cite{MKB,exactsca,jain,RK,ca-mass}. Hence we expect
that the present model is also well described by the mean field
theory. To prove no transitions in the thermodynamic limit, we
first assume the existence of the condensed phase, so that there
is an aggregate. We also assume that all sites except the site
occupied by an aggregate are occupied by mass $\rho_c$, which
reflects the uniform background distribution with mass $\rho_c$.
Then, for the existence of an aggregate in the steady states, the
mass loss of an aggregate by chipping should be equal to the gain
from the background. However the mass of an aggregate diverges
with $L$, so the chipped mass of an aggregate also diverges as
$L^\lambda$. On the other hand, the background distribution should
not change in the condensed phase, the gain from the background is
finite and independent of $L$. As a result, the loss diverges in
the thermodynamic limit while the gain is finite. Therefore an
aggregate cannot exist in the thermodynamic limit, which
contradicts with the assumption of the existence of an aggregate.
So the condensed phase cannot exist and the system is always in
the fluid phase in the limit $L \to \infty$. Since the argument is
based on the mean field, we expect no transitions in any
dimensions.

In summary, we investigate the condensation phase transitions of
SCA model with a mass-dependent fragmentation in one dimension. We
numerically find that condensation transitions take place only in
finite systems. The critical density diverges with system size $L$
as $L^{\beta}$ because of $\tau < 2 $ for $0<\lambda <1$. Hence in
the thermodynamic limit of $L \to \infty$, no condensation
transitions occur, so that a system is always in the fluid phase
for $\lambda > 0$. Based on the mean field argument, we conclude
that there are no condensation transitions in any dimensions.

We thank Prof. S. H. Yook and Dr. S. Y. Yoon for helpful
suggestions and critical comments. This work was supported by the
Korea Science and Engineering Foundation(KOSEF) grant funded by
the Korea government(MOST) (No. R01-2007-000-10910-0).

\end{document}